\documentclass[12pt,preprint]{aastex}

\shorttitle{Transition from L to T Dwarfs }
\shortauthors{T. Tsuji \& T. Nakajima}

\begin{document}

\title{Transition from L to T Dwarfs on the  Color-Magnitude Diagram}

\author{TAKASHI TSUJI\altaffilmark{1}\altaffiltext{1}{Institute of Astronomy, 
School of Science, The University of Tokyo, 2-21-1 Osawa, Mitaka, Tokyo, 
181-0015, Japan; ttsuji@ioa.s.u-tokyo.ac.jp}  and 
TADASHI NAKAJIMA\altaffilmark{2}\altaffiltext{2}{National Astronomical 
Observatory, 2-21-1 Osawa, Mitaka, Tokyo, 181-8588, Japan; 
tadashi.nakajima@nao.ac.jp } }

\begin{abstract}

The color-magnitude (CM) diagram of cool dwarfs  and brown dwarfs based 
on the  recent astrometry data is compared with the CM diagram transformed 
from the theoretical evolutionary tracks via the unified cloudy models 
(UCMs) of L and T dwarfs. A reasonable agreement between the models and 
observations is shown for the whole regime of ultracool dwarfs  covering 
L and T dwarfs, and this is achieved, for the first time, with the use of 
a single grid of self-consistent nongray model photospheres accommodating  
dust cloud (UCMs; $ 700 \la T_{\rm eff} \la 2600$ K). A distinct brightening  
at the $J$ band in the early T dwarfs revealed by 
the recent parallax  measurements is  explained as a natural 
consequence of the migration of the thin dust cloud to the inner region
of the photosphere and should not necessarily be evidence for
Burgasser et al.'s proposition that the dust cloud breaks up in the L/T 
dwarf transition. Also, the rapid bluing from the late L to the early T 
dwarfs is a direct result of the transition of the thin dust cloud from 
the optically thin ($\tau < 1$) to thick ($\tau \ga 1$) regimes while 
$L_{\rm bol}$ and $T_{\rm eff}$  lower only slightly. Thus, 
the theoretical evolutionary models, the cloudy models of the 
photospheres (UCMs), and the observed fundamental stellar parameters are 
brought into a consistent picture of the newly defined L and T dwarfs.

\end{abstract}


\section{INTRODUCTION}

Since the discovery of the genuine brown dwarf Gl\,229B \citep{nak95},
a large number of ultracool dwarfs have been discovered 
\citep[e.g.][]{str99, kir00, bur02a}. The newly discovered 
ultracool dwarfs are now classified into the new spectral classes L and T 
\citep{geb02}, which are characterized by the very red colors due to dust 
extinction and by the strong bands of methane 
and water, respectively. Extensive efforts have been made to interpret 
these new objects in terms of atmospheric chemistry including formation
of dust clouds, as reviewed recently by  Burrows et al. (2001). 
Although L and T dwarfs appear to be quite different, we proposed that
they can be understood consistently with 
the unified cloudy models (UCMs) in which a thin dust cloud is formed 
always near the dust condensation temperature and hence will be located 
relatively deep in the photosphere for the cooler T dwarfs while will appear 
in the optically thin regime in the warmer L dwarfs \citep{tsu02}. The 
presence of the dust cloud deep in the photospheres was also shown by 
an application of the planetary theory \citep{mar02}. It is interesting 
that the presence of the thin dust cloud deep in the photospheres of 
ultracool dwarfs has been concluded from the quite different approaches.

The nature of the cloud, however, is by no means clear yet. 
One major interest is if the cloud is  subject to the meteorological 
activities familiar with the planets in the solar system. Such a
possibility has so far been suggested by observing photometric variabilities
which may be due to the inhomogeneity of the dust clouds \citep[e.g.][]
{bai01,mart01}. 
Also, an interesting finding of the recent parallax measurements
is that the $J$ magnitude shows a large brightening in the 
early T dwarfs \citep{dah02}. This result was interpreted as 
evidence for the disruption of the dust cloud in  the L/T dwarf transition
\citep{bur02b}.  

Before such a possibility is explored, however,  it is important to 
remember a more classical application of the CM diagram; namely as a 
touchstone of stellar models. For this purpose, the observed CM 
diagram of ultracool dwarfs including  brown dwarfs is now accurate 
enough to be confronted with the theoretical evolutionary models
\citep[e.g.][]{dah02}. 
On the other hand, theoretical evolutionary tracks of the substellar
mass objects have been discussed by \citet{hay63} already in the 1960's, and 
recent developments with improved input physics have been discussed by  
\citet{burr97} and by \citet{cha00}. Given that the observed CM diagram and 
theoretical evolutionary models are both reasonably well established,
a missing link between them is a realistic  model photosphere
to be used for the conversion of the fundamental stellar parameters to
the observables.  We show in this letter that the missing link
might be found in the UCMs (Tsuji 2002).
  
\section{COLOR-MAGNITUDE DIAGRAM }

We applied the UCMs to convert the $M_{\rm bol}$ and $T_{\rm eff}$
of the evolutionary models to more easily observable monochromatic absolute 
magnitudes and color indices; we discuss  $M_J$ and $J-K$  as an example. 
In the UCMs, we assume that the dust grains formed at the condensation 
temperature ($T_{\rm cond}$) are  in detailed balance with the ambient 
gaseous mixture so long as the grain sizes are smaller than the critical 
radius ($r_{\rm cr}$) and hence can be sustained in the photosphere. Dust 
grains soon grow to be larger than $r_{\rm cr}$ at a slightly lower 
temperature which we refer to as the critical temperature ($T_{\rm cr}$) 
and the dust grains are stabilized. Such stable grains may segregate from 
the gaseous mixture and can no longer be sustained in the photosphere. 
The exact value of $r_{\rm cr}$ is difficult to know but we assume it to 
be quite small in the sub-micron regime (e.g. $r_{\rm cr} \approx 0.01\mu$m), 
since astronomical grains prevailing in the universe (and hence these 
grains are already larger than $r_{\rm cr}$) are well above this size as 
can be inferred from the reddening law. For this reason,
only  dust grains smaller than  $r_{\rm cr}$ survive in the temperature range 
of $ T_{\rm cr} \la T \la T_{\rm cond} $. This means  a formation of a dust 
cloud with a rather high temperature (note that $T_{\rm cond} \approx 
2000$\,K) independently of $T_{\rm eff}$. Since $ T \approx 
T_{\rm eff}$ at $\tau_{\rm Ross} \approx 1$, the 
dust cloud appears in the optically thin region ($\tau_{\rm Ross} < 1$)
in L dwarfs whose $ T_{\rm eff}$'s are relatively high and in the 
optically thick region ($\tau_{\rm Ross} \ga 1$) in T dwarfs whose 
$ T_{\rm eff}$'s are lower. As a result, the thin dust cloud formed
in the photosphere of ultracool dwarfs moves from the optically thin
region in L dwarfs to the deeper optically thick region in T dwarfs.
This migration of the dust cloud in the photospheres of ultracool dwarfs 
has observable effects to be reflected on the observed CM diagram, 
as we will  show below. 

In discussing model photosphere, one important input parameter is
the chemical composition. After the first version of the UCMs \citep{tsu02},
an important revision of the solar carbon and oxygen abundances was 
proposed by Allende Prieto, Lambert, \& Asplund (2001, 2002), who showed 
that the C and O abundances are about 50\% lower than those by \citet{and89}. 
In view of the decisive role of C and O in the chemical equilibrium in
cool dwarfs \citep{lod02},
we have computed a new grid of the UCMs with the new C 
and O abundances. The other abundances are by \citet{and89} and we assumed
that the abundant refractory elements Fe, Si, Mg, and Al are
condensed into iron, enstatite (MgSiO$_3$), and corundum (Al$_2$O$_3$). 
The extinction coefficients
of these dust species are evaluated with the use of the measured optical
constants and they do not depend on the grain size so far as the grains
are as small as we have assumed (Tsuji 2002).
   
To convert $M_{\rm bol}$ predicted from the evolutionary models to 
$M_{\rm J} $, we apply  the   bolometric correction for the $J$ magnitude 
defined by
     $$ BC_{\rm J} = M_{\rm bol} - M_{\rm J}, \eqno(1) $$
with     
$$ M_{\rm bol}  = -2.5 {\rm log} \int_0^{\infty} { {4\pi R^2 F_{\lambda}}\over 
           {4 \pi d_{10}^2} } d\lambda + C_1, \eqno(2) $$
and 
$$   M_{\rm J}  =  -2.5 {\rm log} \int_{\lambda_1}^{\lambda_2}{ {4\pi R^2 
S_{\lambda}^{J} F_{\lambda} } \over {4 \pi d_{10}}^2} d\lambda + C_2, 
                                              \eqno(3)  $$
where $F_{\lambda}$ is the emergent flux calculated from our UCM with the 
use of the molecular linelist,
$ C_1 $ and $ C_2 $ are constants which depend on the zero-points
of the magnitude systems, $R$ is the stellar radius, and $d_{10}$ is the
distance to the object corresponding to  10\,pc. 
We used the filter response function  $S_{\lambda}^{J}$ given by 
\citet{per98}, which is normalized as
$$ \int_{\lambda_1}^{\lambda_2} S_{\lambda}^{J} d\lambda = 1.0. 
                                                    \eqno(4) $$
Then
 $$  BC_{\rm J}=-2.5 \biggl[  {\rm log}  \int_0^{\infty} F_{\lambda}d\lambda
- {\rm log} \int_{\lambda_1}^{\lambda_2} S_{\lambda}^{J} F_{\lambda}d\lambda 
  \biggr]+ C,                                             \eqno(5)     $$
where $C = C_1 - C_2$.
We first apply eqn.(5) to determine $C$ by the use of the model flux of 
Vega with $ T_{\rm eff} = 9550$\,K, log $g  = 3.95$, and $ V_{\rm micro} =
2$\,km s$^{-1}$ \citep{kur93}. The value of $BC_{\rm J}$ for Vega is $-0.22$ 
from $ m_{\rm bol} = -0.22$ \citep{cod76}  and
$ J = 0.0 $. We found $ C = -7.72 $, with which eqn.(5) is applied
to the model fluxes of L and T dwarfs. 

The resulting values of  $BC_{\rm J}$ against $T_{\rm eff}$ are shown 
in Fig.\,1a for the four values of $T_{\rm cr}$, the fully dusty models 
(case B; $T_{\rm cr} = T_{\rm surface}$), dust segregated 
models (case C; $T_{\rm cr} = T_{\rm cond}$ and hence effectively 
dust-free), and the blackbody (BB) radiation. 
We assumed log\,$g = 5.0$ which is the median value of  log\,$g$ 
extending from 4.5 to 5.5 in brown dwarfs (Fig.\,9 of Burrows et al. 1997).
The changes of log\,$g$  by $\pm 0.5$     
at $T_{\rm eff} = 1600$\,K ($T_{\rm cr} = 1800$\,K), for example,
result in the changes of $BC_{\rm J}$ by about $\pm 0.15$ (Fig.\,1a).  
The $BC_{\rm J}$ can be regarded as a generalized color index and 
the larger $BC_{\rm J}$ of 
the case C compared with the case of BB for the lower $ T_{\rm eff} $ is 
due to the large nongray opacity which is relatively transparent in the 
$J$ band region compared to the longer wavelength region dominated by the 
increasingly larger absorption due to H$_2$ CIA (collision-induced absorption)
as well as CH$_4$ and H$_2$O ro-vibration bands.  
If the thin cloud is introduced, 
the dust extinction at the $J$ band region first increases from the early L 
to late L dwarfs and $BC_{\rm J}$ decreases. 
The dust column density in the observable photosphere attains the maximum
value of $7.5 \times 10^{-3}$ gram cm$^{-2}$ at $T_{\rm eff} \approx
1500$\,K which corresponds to the late L dwarfs.
In T dwarfs, however, the cloud moves into the optically thick 
region of the photosphere, and the observable effect of the dust extinction 
decreases in the L/T transition objects and further in  T dwarfs, 
resulting in the re-increase of $BC_{\rm J}$. 
The values of $BC_{\rm J}$ depend on  $ T_{\rm cr}$, which is essentially 
a measure of the thickness of the dust cloud in the observable
photosphere $(\tau \la 1)$; the lower value of 
$ T_{\rm cr}$ implies the thicker cloud and hence a larger dust 
extinction, resulting in the smaller values of $BC_{\rm J}$. For the case B, 
the dust extinction is so large that the $J$ flux is suppressed below the 
case of BB and hence the values of $BC_{\rm J}$  tend to be
smaller than those for the case of BB at $ T_{\rm eff} \la 1700$\,K.

Next, we evaluate $J - K$ based on the predicted fluxes of the UCMs, 
and the resulting  $J-K$ is reduced to the empirical system so 
that  $J - K = 0.0$ for A0V star, again by the use of the model flux of 
Vega by \citet{kur93}. We used the filter response function of \citet{per98} 
and their transform equation to the CIT system. The resulting $T_{\rm eff} 
- (J-K)$ calibrations are shown in Fig.\,1b for the four values of  
$T_{\rm cr}$, the cases B and C, and the case of BB. The red limit of 
$J-K$ depends on $T_{\rm cr}$ and is larger (redder) for the lower value of 
$T_{\rm cr}$, which implies the larger dust column density of the
cloud in the photosphere; 
$(J-K)_{\rm max} \approx 2.3, 2.0, 1.7$ \& 1.4 for $T_{\rm cr} = 1700, 
1800, 1850,$ \& 1900\,K, respectively (Fig.\,1b). On the other hand, the 
observed red limit of $J-K$ is about 1.9 from  Fig.\,4 of \citet{dah02}. Thus 
we suggest that $T_{\rm cr} \approx 1800$\,K in agreement with the previous 
analyses based on the different color systems (Tsuji 2001, 2002). 
This result is based on the grid of UCMs with log\,$g = 5.0$, the changes of
which by $\Delta\,{\rm log}\,g = \pm0.5$ result in $\Delta\,(J-K) \approx 
\mp0.2$ (Fig.\,1b), and this may partly 
explain the scatter in the observed colors (Fig.\,4 of Dahn et al. 2002).

We now transform  $M_{\rm bol}$  and  $T_{\rm eff}$  of the evolutionary 
models  to $M_{\rm J}$ and $J - K$ via Figs.\,1a and 1b with $T_{\rm cr} 
= 1800$\,K throughout. We first apply this procedure to the cooling 
tracks of the brown dwarfs with 10, 42, and 70\,$M_{\rm Jupiter}$ by 
\citet{burr97}, and the results are shown in Fig.\,2a  with the observed 
$ M_{J}$ and $J-K$ values \citep{dah02}. It is to be noted that the red 
limit of L dwarfs as well as the rapid bluing of T dwarfs is well 
reproduced by the cooling tracks of the substellar mass objects. 
The position of the T2 dwarf SDSS\,J1254-0122 is well accounted 
for by the  cooling track of the moderate mass  brown dwarf and this fact 
confirms that this object is in fact a transition object from L to T dwarf 
\citep{leg00}. The high luminosity of the T5 dwarf 2MASS\,J0559-1404 at 
the $J$ band  is also accounted for by the cooling track of the low 
mass brown dwarfs. A few objects near $M_{J} \approx 15$  may be 
accounted for by the models of higher masses while the observed 
photometric data of the very faint Gl\,570D may not be well fixed yet. 

Next, we repeat the same analysis on the theoretical isochrones for 
0.1, 1 and 10\,Gyr by \citet{cha00} and the results are shown in Fig.\,2b 
with the observed data.  The observed position of 
2M\,0559 is still higher than the isochrone of 0.1\,Gyr, but a possibility 
of a binary was suggested \citep{bur02} and then the observed position 
may be shifted downward by 0.75\,mag.  The results shown 
in Figs.\,2a and 2b are not fully self-consistent in that the model 
photospheres used in computing the evolutionary models are not 
the same as those used in converting the resulting $(M_{\rm bol}, 
T_{\rm eff})$ diagrams to the CM diagrams. According to \citet{burr97},
however, by replacing gray models with nongray  models, the evolutionary 
tracks have changed, but generally by no more than 10\% in luminosity at 
any time, for any given mass. Also, the effect of the photospheric models 
on  $T_{\rm eff}$ was shown to be no larger than about 100\,K (Fig.\,2 of 
Chabrier et al. 2000). With these reservations in mind, we conclude that the 
general trend of the observed CM diagram can be reasonably well fitted with 
the predicted ones based on the evolutionary models and the photospheric 
models (and Figs.\,1a \& 1b can also be applied to any other evolutionary
track).

It is to be noted that the ``brightening'' of the $J$ flux in the early 
T dwarfs is an artifact of observing objects with different masses and/or 
ages, and it should not imply that a single cooling track of a given mass  
shows the  ``brightening''. Nevertheless the rather
high luminosity at the $J$ band reflects the re-increase of $BC_{\rm J}$ 
at the L/T transition noted before and this  is a natural consequence of 
the migration of the dust cloud from the optically thin region $(\tau < 1)$ 
in L dwarfs to the optically thick region $(\tau \ga 1)$  in T dwarfs. 
This migration of the dust cloud  recovers the  gaseous opacities  
 including H$_2$ CIA at the $K$ band region and makes the
$J$ flux large enough to compensate for the decreasing bolometric 
flux in the early T dwarfs. Also, the rapid transition  from  L to T dwarfs 
on the CM diagram is simply because the cloud moves from the optically 
thin to thick regions for only a small change of  $T_{\rm eff}$ during the
L/T transition. It is to be emphasized here that the migration  of 
the dust cloud to the deeper photosphere in the cooler brown dwarfs is 
simply a  consequence of the dust formation at the condensation temperature 
($T_{\rm cond}\approx 2000$\,K) irrespective of $T_{\rm eff}$, and no 
other mechanism is needed at all to explain the migration of the dust cloud. 
 
\section{DISCUSSION AND CONCLUDING REMARKS}

  We have shown that the observed CM diagram extending from  L to T
dwarfs can be well fitted with the theoretical $(M_{\rm bol}, T_{\rm eff})$ 
diagram converted to ($M_{J}, J-K)$ diagram via UCMs. This is the first 
successful application of the model photospheres to the CM diagram of the 
ultracool dwarfs which show rapid bluing and unexpected ``brightening'' 
at the $J$ magnitude in T dwarfs after passing the red limit in  L dwarfs.
So far, other model photospheres including the cloudy models by \citet{mar02} 
as well as the dusty and clear models by \citet{cha00} could not be 
fitted to the observed CM diagram (Fig.\,1 of Burgasser et al. 2002b). 
Although the cloudy models of \citet{mar02} have essentially the same 
feature as our UCMs in that the thin dust cloud is formed deep in the 
photosphere, yet their models  failed to account for the observed $(M_{J}, 
J-K)$ diagram, even if their free parameter referred to as the 
sedimentation efficiency, $f_{\rm rain}$, was changed. 

The difficulty of the Marley et al's cloudy models led \citet{bur02b}
to suggest that the dust cloud should subject to disruption. 
Then the breaks between the clouds could explain the brightening
of $M_{J}$ as well as the bluing of $J-K$ of the T5 dwarf 2M\,0559,
but only by introducing an additional free parameter, the cloud coverage 
fraction, besides the sedimentation efficiency, $f_{\rm rain}$.    
Also, the position of the L/T transition object SDSS\,1254 on the
CM diagram could be explained only if the cloud coverage of 40\% was
assumed. However, the argument based on the
failure of the Marley et al's cloudy models to explain the  observed CM 
diagram cannot be the evidence for the dust cloud disruption, since our 
simpler cloudy models, the UCMs, easily explain the observed $(M_{J}, J-K)$
diagram including the key objects SDSS\,1254 and 2M\,0599 
without any additional ad-hoc assumption, as shown in Sect.2.
As to an additional argument for the cloud disruption, namely the 
non-monotonic behavior of the FeH $F ^{4}\Delta - X ^{4}\Delta$ 
(0 - 0) band at 0.9896 $\mu$m \citep{bur02b}, further observational
confirmation  will be needed since no FeH (1 - 0) band at 0.8692 $\mu$m 
can be seen in the spectrum of the T5 dwarf 2M\,0559 (Fig.\,6.4 of Burgasser 
2002).

In conclusion, the observed CM diagram based on 
the recent parallax measurements can be fitted with the predicted CM 
diagrams based on the evolutionary models and UCMs of the photospheres. 
Since the theoretical evolutionary models and observed CM diagram
can now be deemed as reasonably well established, this success in
fitting the theoretical and observed CM diagrams can be regarded as
an observational confirmation of the model photospheres we have
applied, namely the UCMs. This success of the UCMs may be significant
especially because the UCMs are based on the very simple thermodynamical
argument that the dust formed at  $T \approx T_{\rm cond}$ can survive
only to where $T \approx T_{\rm cr}$, resulting in the formation of a thin 
dust cloud  between $T \approx T_{\rm cr}$ and $T_{\rm cond}$. Then, 
possibly complicated processes including condensation, growth, segregation, 
and precipitation of dust grains in the cloud are all absorbed in the 
$T_{\rm cr}$, which is determined empirically. 
Such a semi-empirical approach plays an important role 
in stellar modeling, and a well known example of a useful empirical 
parameter is the so-called mixing length  in the theory of stellar 
convection. Now our simplified  
treatment of the dust cloud has been shown to be successful  
for the interpretation of the CM diagram and will hopefully be useful for 
the interpretation and analysis of other observed data as well. 
\vspace{-5mm}

\acknowledgements
We thank the anonymous referee for a careful reading and helpful
comments. This work was  supported by the grants-in-aid of JSPS Nos.11640227 
(T.T.) and 14520232 (T.N.).

\clearpage

\begin{figure}
\epsscale{0.6}
\plotone{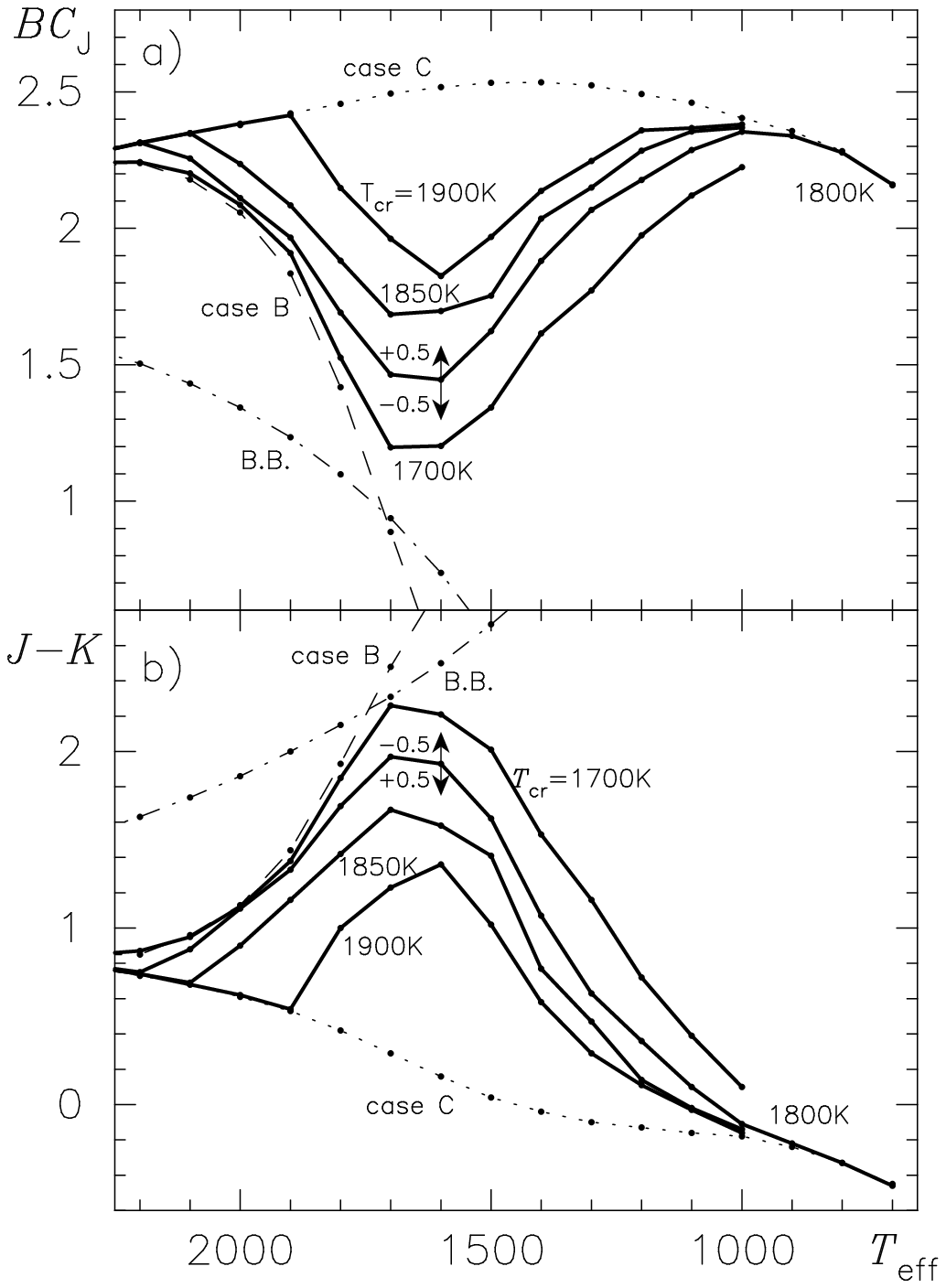}
\caption {
a) The values of $BC_{\rm J}$ based on the unified cloudy models (UCMs) of
log\,$g = 5.0$ with $T_{\rm cr} = 1700, 1800, 
1850,$ and 1900\,K are plotted against 
$T_{\rm eff}$ by the solid lines. Also, those for the fully
dusty models (case B), dust segregated models (case C), and blackbody
radiation are shown by the dashed, dotted, and dash-dotted lines,
respectively. The effects of changing log\,$g$ by $\pm 0.5$ for the 
case of $T_{\rm eff} = 1600$\,K and $T_{\rm cr} = 1800$\,K are shown by 
the vectors.
b) The same as for a) but for the values of $J-K$. Note that the dust cloud
results in essentially the same effect on $BC_{\rm J}$ and on $J-K$
(apart from the sign).
}
\label{Fig1}
\end{figure}

\clearpage

\begin{figure}
\epsscale{0.9}
\plotone{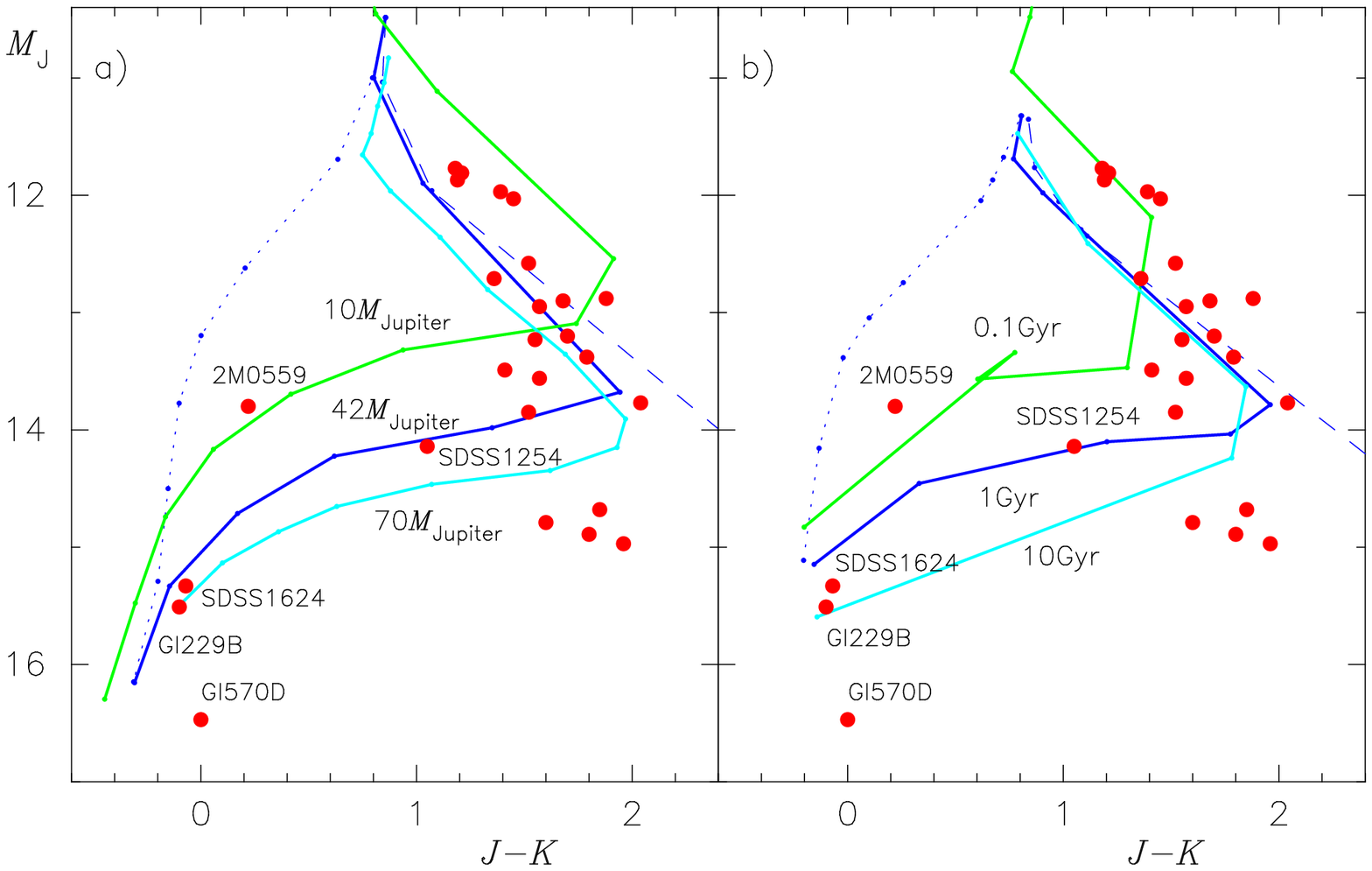}
\caption {
a) The cooling tracks of brown dwarfs with $M = 10, 42,$ and 
$70\,M_{\rm Jupiter}$ by \citet{burr97} converted to CM diagrams via 
unified cloudy models (UCMs) of $T_{\rm cr} = 1800$\,K (solid lines) are
compared with  the observed data (filled circles) by \citet{dah02}.  
Also, shown are those with $M = 42 M_{\rm Jupiter}$ converted via the fully
dusty models of case B (dashed line) and via the dust segregated models of 
case C (dotted lines). 
b) The same as for a) but for the isochrones of brown dwarfs for 0.1, 1.0 
and 10\,Gyr by \citet{cha00} converted to CM diagrams via UCMs of 
$T_{\rm cr} = 1800$\,K (solid lines).
Also, those for 1\,Gyr isochrone converted via the models of cases B 
(dashed line) and  C (dotted lines) are shown.
}
\label{Fig2}
\end{figure}


\begin{thebibliography}

\bibitem[Allende Prieto et al.(2001)]{alle01}
Allende Prieto, C., Lambert, D. L. \& Asplund, M. 2001, \apj, 556, L63 

\bibitem[Allende Prieto et al.(2002)]{alle02}
Allende Prieto, C., Lambert, D. L. \& Asplund, M. 2002, \apj, 573, L137 

\bibitem[Anders \& Grevesse(1989)]{and89}
Anders, E., \& Grevesse, N. 1989, Geochim. Cosmochim. Acta, 53, 197 

\bibitem[Bailer-Jones \& Mundt(2001)]{bai01} 
Bailer-Jones, C. A. L., \& Mundt, R. 2001, \aap, 367, 218 

\bibitem[Burgasser (2002)]{bur02} 
Burgasser, A. J. 2002, Ph. D. thesis, California Institute of Technology
  
\bibitem[Burgasser et al.(2002a)]{bur02a} 
Burgasser, A. J., et al. 2002a, \apj, 564, 421 

\bibitem[Burgasser et al.(2002b)]{bur02b} 
Burgasser, A. J., et al. 2002b, \apj, 571, L151 

\bibitem[Burrows et al.(2001)]{burr01} 
Burrows, A., Hubbard, W. B., Lunine, J. I., \& Liebert, J. 2001, 
Rev. Mod. Phys., 73, 719


\bibitem[Burrows et al.(1997)]{burr97} 
Burrows, A., et al. 1997, \apj, 491, 856

\bibitem[Chabrier et al.(2000)]{cha00} 
Chabrier, G., Baraffe, I., Allard, F.,  \& Hauschildt, P. 2000, \apj,
542, 464

\bibitem[Code  et al.(1976)]{cod76} 
Code, A. D., Davis, J., Bless, R. C., \& Hanbury Brown, R. 1976, \apj, 203, 
417

\bibitem[Dahn  et al.(2002)]{dah02} 
Dahn, C. C., et al. 2002, \aj, 124, 1170

\bibitem[Geballe  et al.(2002)]{geb02} 
Geballe, T. R., et al. 2002, \apj, 564, 466

\bibitem[Hayashi \& Nakano(1963)]{hay63}
Hayashi, C., \& Nakano, T. 1963, Prog. Theor. Phys., 30, 460 

\bibitem[Kirkpatrick et al.(2000)]{kir00}
Kirkpatrick, J. D., et. al. 2000, \aj, 120, 447

\bibitem[Kurucz(1993)]{kur93} 
Kurucz, R. L. 1993,  Kurucz CD-ROM 13, ATLAS9 Stellar Atmosphere
Programs and 2 km/s Grid (Cambridge: SAO)

\bibitem[Leggett et al.(2000)]{leg00} 
Leggett, S. K., et al. 2000 \apj, 536, L35

\bibitem[Lodders \& Fegley(2002)]{lod02} 
Lodders, K., \& Fegley, B., Jr. 2002, Icarus, 155, 393

\bibitem[Marley et al.(2002)]{mar02} 
Marley, M. S., Seager, S., Saumon, D., Lodders, K., Ackerman, A. S., 
Freedman, R., \& Fan, X. 2002, \apj, 568, 335

\bibitem[Mart\'in, Zapatero Osorio, \& Lehto(2001)]{mart01} 
Mart\'in, E, L., Zapatero Osorio, M. R., \& Lehto, H. J. 2001, \apj, 557, 822

\bibitem[Nakajima et al.(1995)]{nak95}
Nakajima, T., Oppenheimer, B. R., Kulkarni, S. R., Golimowski, D. A., 
Matthews, K., \& Durrance, S. T. 1995, \nat, 378, 463 

\bibitem[Persson et al.(1998)]{per98}
Persson, S. E., Murphy, D. C., Krzeminski, W., Roth, M., \& Rieke, M. J.
1998, \aj, 116, 2475

\bibitem[Strauss et al.(1999)]{str99} 
Strauss, M. A., et al. 1999, \apj, 522, L61

\bibitem[Tsuji(2001)]{tsu01}
Tsuji, T. 2001, in Ultracool Dwarfs:  New Spectral Types L and T,  
ed. H. R. A. Jones \& I. A. Steele (Berlin: Springer-Verlag), 9

\bibitem[Tsuji (2002)]{tsu02}
Tsuji, T. 2002, \apj, 575, 264 

\end{thebibliography}
\end{document}